\documentclass[a4paper,10pt]{article}

\usepackage{tensor}
\usepackage{amsmath}
\usepackage{amsfonts}
\usepackage{amssymb}
\usepackage{pstricks}
\usepackage{graphicx}
\usepackage{epsfig}

\title{Fluctuations of spacetime and holographic noise in atomic interferometry}
\author{Ertan G\"okl\"u and Claus L\"ammerzahl\\ 
ZARM, University Bremen, Am Fallturm, 28359 Bremen, Germany}

\begin{document}

\maketitle

\begin{abstract}
Space--time can be understood as some kind of space--time foam of fluctuating bubbles or loops which are expected to be an outcome of a theory of quantum gravity. One recently discussed model for this kind of space--time fluctuations is the holographic principle which allows to deduce the structure of these fluctuations. We review and discuss two scenarios which rely on the holographic principle leading to holographic noise. One scenario leads to fluctuations of the space--time metric affecting the dynamics of quantum systems: (i) an apparent violation of the equivalence principle, (ii) a modification of the spreading of wave packets, and (iii) a loss of quantum coherence. All these effects can be tested with cold atoms. These tests would supplement measurements of a so called ``mystery noise'' at the gravitational wave detector GEO600 which was recently speculated to have its origin in holographic noise.

\end{abstract}

\section{Introduction}

The unification of quantum mechanics and General Relativity is one of the most outstanding problems of contemporary physics. Though yet there is no theory of quantum gravity. There are a lot of reasons why gravity must be quantized \cite{Kiefer04}. For example, (i) because of the fact that all matter fields are quantized the interactions, generated by theses fields, must also be quantized. (ii) In quantum mechanics and General Relativity the notion of time is completely different. (iii) Generally, in General Relativity singularities appear necessarily where physical laws are not valid any more. Such singularities probably can be avoided by a theory of Quantum Gravity. 

Approaches like string theory or loop quantum gravity make some general prediction which may serve as guidance for experimental search. The difficulty is that in many cases the precise strength of the various expected effects are not known. 
One class of expected effects of a quantized theory of gravity is related to a nontrivial quantum gravity vacuum (depicted by Wheeler as "space--time foam'') which can be regarded as a fluctuating space--time. 

Space--time fluctuations could lead to a minimal observable distance setting an absolute bound on the accuracy of the measurement of distances and defining a fundamental length scale \cite{{JaekRey1994PysLettA}, {KlinhamArxiv}}. For instance,  corresponding additional noise sources in gravitational wave interferometers has been  considered \cite{GAC2000PRD}, which has also been analyzed in the context of an experiment with optical cavities \cite{SchillPRD2004}. A space--time foam may also violate the principle of equivalence as was suggested by Ellis and coworkers \cite{EllMavNanSak2004}. Another prediction of quantum gravity models are so called deformed dispersion relations \cite{DeformedDispers0}-\cite{DeformedDispers3}.
In the work of Hu and Verdauger \cite{HuVerd2004LivRev} classical stochastic fluctuations of the space--time geometry were analyzed stemming from quantum fluctuations of matter fields in the context of a semiclassical theory of gravity. This leads to a stochastic behavior of the metric tensor. Furthermore, the effects of fluctuations of the space--time geometry which lead to, e.g., lightcone fluctuations, redshift, and angular blurring were discussed in \cite{FordIntJTP2005}.
Space--time fluctuations can also lead to the decoherence of matter waves which was discussed in \cite{PowerPercival1999}, \cite{Wang2006CQG}, \cite{BonifacioWangMendocaBingham09} and \cite{BreuerGoeklueLaemml09}.
The analysis in \cite{Camacho2003GRG} takes into account nonconformal fluctuations of the metric and yields a modified inertial mass which is subject to the stochastic properties while - generalizing this work - in \cite{GoeklueLaemmerzahl08} the implications on a modified mass and an apparent violation of the weak equivalence principle were analyzed.

In this paper we will lay special emphasis on the implications of a holographic noise. Recently, C. Hogan presented a special model of holographic noise and claimed that this could explain the so-called ``GEO600 mystery noise''. In the next sections we will discuss how the holographic principle can lead to space--time fluctuations and, thus, whether the observed mystery noise in GEO600 could have its origin in quantum gravity.  

The paper is organized as follows: We start with an overview of the holographic principle and present two scenarios leading to holographic noise. The first scenario leading to metrical fluctuations has been extensively discussed by Ng and coworkers \cite{JackNg2003},\cite{JackNg2002}, and the second is devoted to the work of Hogan \cite{Hogan_01_09}, where the holographic noise manifests as transversal fluctuations of the position of the beam splitter in GEO600 (transversal with respect to the direction of the laser beam). Hereafter, we will discuss how fluctuations of the metric modify the dynamics of quantum systems. We will show that this leads to several modifications: (i) the inertial mass of quantum particles is modified accompanied by an apparent violation of the weak equivalence principle. (ii) The spreading of wave packets is affected by space--time fluctuations and (iii) quantum particles are subject to decoherence. The discussion concerning the type of metrical fluctuations (noise scenario) is as general as possible and we will also show how holographic noise affects all aforementioned modifications of quantum dynamics. In all cases we discuss related interferometry experiments with cold atoms.

\section{Models of holographic noise}\label{MODHOL}

In the following we shortly review how the notion of ``holographic'' noise appears in the physics of space--time in an operational sense and introduce two different models. The first describes metric fluctuations and the second acts effectively like a shear of the metric. 

\subsection{Bounds on length measurements}

We start with the question how precisely we can measure a distance $l$ when quantum mechanics is taken into account. This problem was introduced by Salecker and Wigner \cite{SaleckerWigner1958} who carried through a {\it gedankenexperiment} and showed that length measurements are bounded by the mass of the measuring device. The measuring device is composed of a clock, a photon detector, and a source (e.g. laser), which altogether possesses a mass $m$ and a typical size $a$. A mirror is placed at a distance $l$ where the light coming from the source is reflected.  Then it turns out that by virtue of the quantum mechanical uncertainty relation the uncertainty of the detector velocity $\Delta v$ is related to the position uncertainty $\Delta x_1$ of the detector according to
\begin{equation} 
\Delta v=\frac{\hbar}{2m\Delta x_1} \, .
\end{equation}
After the light round trip time $T=2l/c$ the detector moved by $T \Delta v$ which yields for the total position uncertainty of the detector, by adding up linearly,  
\begin{eqnarray}
\Delta x_{\rm tot} = \Delta x_1 + T \Delta v = \Delta x_1 + \frac{\hbar T}{2 m \Delta x_1} \, .
\end{eqnarray}
Minimizing this expression w.r.t. $\Delta x_1$ yields the minimum position uncertainty of the detector given by
\begin{eqnarray}
\delta l_{\rm qm}=(\Delta x_{\rm tot})_{\rm min}=2 \sqrt{\frac{\hbar l}{mc}} \, , \label{SaleckerWigner}
\end{eqnarray}
which is the result of the famous \textit{Salecker and Wigner Gedankenexperiment} \cite{SaleckerWigner1958}. This is also the so--called \textit{standard quantum limit} which is important for the precision of measurements in gravitational wave detectors. Therefore, in principle, the error in the position measurement can be reduced arbitrarily by increasing the mass $m$. However, this is limited by the requirement that the detector should not turn into a black hole. 
This condition calls for the inclusion of gravity in the analysis of the measurement uncertainty 

From the requirement that the detector must not be a black hole we set $l > a > r_{\rm S}$,  where $r_{\rm S}$ denotes the Schwarzschild radius, and one can suspect a first estimate for the total measurement uncertainty. By inserting $l\gtrsim r_{\rm S}$ we get  
\begin{equation}
\delta l_{\rm qm} \gtrsim l_{\rm P}, \label{naiveestimate}
\end{equation}
clearly indicating that the uncertainty is limited by the Planck length $l_{\rm P}=\sqrt{G\hbar/c^3}$ ($\approx 10^{-35}$m). 

However, in the following we will see that the uncertainty is larger than in the original Salecker Wigner case when curvature effects are taken into account. 
We consider a spherically symmetric detector surrounded by a Schwarzschild metric, then the optical path from the light traveling from the detector (starting at the position $a> r_{\rm S}$) to a point $l>a$ is the well known result
\begin{equation}
 c \Delta t = \int^{l}_{a} \frac{d\rho}{1-r_s/\rho} = (l-a)+ r_s \log{\frac{l-r_s}{a-r_{\rm S}}},
\end{equation} 
which can be approximated for $l > a \gg r_{\rm S}$ leading to the gravitational contribution of the error of the length measurement 
\begin{equation}
\delta l_{\rm g} = r_{\rm S} \log{\frac{l-r_{\rm S}}{a-r_{\rm S}}} \approx r_{\rm S} \log{\frac{l}{a}} \, .
\end{equation}
If we measure a distance $l\geq 2a$ then $\delta l_{\rm g} \geq  \frac{Gm}{c^2}$ and the {\it total error} reads 
\begin{equation}
\delta l_{\rm tot} = \delta l_{\rm qm} + \delta l_{\rm g} = 2 \sqrt{\frac{\hbar l}{mc}} + \frac{Gm}{c^2} \, .
\end{equation}
By minimizing the total error w.r.t. the mass $m$ of the detector this yields the minimal uncertainty
\begin{equation}
(\delta l_{\rm tot})_{\rm min}= 3 (l^2_p l)^{1/3} \, . \label{holo1}
\end{equation}
In contrast to the first naive estimate (\ref{naiveestimate}) where the measurement uncertainty was bound linearly by the Planck length, we see that the total error is now larger due to the cubic root. Because the Planck length is so short the uncertainty in the distance is still very small. For example, considering the whole universe ($l \approx 10^{10}$ light years) leads to a tiny effect of $(\delta l_{\rm tot})_{\rm min}\approx 10^{-15}$ m.

\subsection{Connection to the holographic principle}

The holographic principle was first proposed by Gerard 't Hooft \cite{tHooftHolo1993} and was later given a string-theory interpretation by Leonard Susskind \cite{SusskindHolo1994}. A good overview is given in \cite{BoussoHolo2002}.
The holographic principle states that the number of degrees of freedom of a volume in space--time is related by the area enclosing the volume in Planck units. Alternatively we can state that although the world appears to have three spatial dimensions, its contents in terms of its degree of freedom can be encoded on a two--dimensional surface, like a hologram. This result comes from black hole physics and the derivation can be sketched as follows. 

Consider a system with entropy $S_0$ inside a spherical region $\Gamma$ bounded by a surface with area $A$, where its mass must be less than that for a black hole. Imagine a spherically symmetric shell of matter collapsing onto the system with the right amount of energy so that it forms together with the system a black hole filling the region $\Gamma$. The entropy is now given by $S=A k_B/(4 l^2_p)$, where $k_B$ is the Boltzmann constant, and obeys the  second law of thermodynamics $dS > 0$. Therefore $S_0 \leqslant S$ and hence $S_0 \leqslant  A k_B/(4l^2_p)$ . This leads to the conclusion that the maximum entropy of a region of space is bounded by its surface area in Planck units. 

After these preparatory remarks we are now in the position to derive the measurement uncertainty $\delta l$ in an alternative way by means of the holographic principle, as was done by Ng and coworkers \cite{JackNg2003,{JackNg2002}}. Suppose we have a spatial cubic region with volume $V=l^3$ which is partitioned into small cubes. The smallest possible cube can be as small as $\delta l^3 $ given by the measurement accuracy $\delta l$. We assume that one degree of freedom (which we will identify with information) is encoded in one such cube. The question is how many degrees of freedom $N$ can be put into the cubic region. According to the holographic principle the maximum amount of information is bounded by the surface area $l^2$ of the cube in units of Planck-area 
\begin{equation}
N= \frac{l^3}{\delta l^3}\leq \frac{l^2}{ l_p^2}
\end{equation}
and we get for the uncertainty $\delta l$
\begin{equation}
\delta l \geq (l_p^2 l)^{1/3}, \label{holo2}
\end{equation}
which is the same result like that obtained by the Gedankenexperiment combining Salecker-Wigner experiment and Gravity (\ref{holo1}). 
The consistency of both results (\ref{holo1}) and (\ref{holo2}) for the uncertainties in length measurements gives us a hint that a microscopic description of space--time may lead to modifications of geometry. In such a quantum description of geometry it is expected that space--time has a foamy structure leading to space--time fluctuations. Since we included gravity in the extended Salecker--Wigner Gedankenexperiment we can suppose that the holographic noise is a possible low--energy remnant of a quantum description of space--time. In this limit the fluctuations can be modeled by fluctuations of the metric so that the fluctuations of a length measurement $\delta l$ translate to metrical fluctuations according to
\begin{equation}
 \delta g_{\mu\nu} \geq (l_p / l)^{2/3} a_{\mu\nu},
\end{equation}
where $a_{\mu\nu}$ is a tensor of order $\mathcal{O}(1)$.

\subsection{Holographic model by Hogan}

In the work of Hogan a  hypothesis is applied about macroscopic position states by considering the results of length measurements being given as position eigenstates of a wavefunction of spacetime \cite{Hogan_01_09}. So spacetime itself is regarded as a frequency-limited wavefunction (limited by the Planck-frequency), or to be more precise, it is the wavefunction of the position of massive bodies within the spacetime, relative to a classical trajectory (which could be a flat metric).
In terms of frames one defines an observer frame including a choice of direction by writing a spacetime wavefunction relative to a \textit{Planck-frequency plane carrier wave}.
In effect this leads to an indeterminacy of relative transverse positions in spacetime. Two measurements of positions of two test masses in a xy-plane (which should have at least masses at the order of the Planck mass) at two different times do not commute. Therefore the relative transverse positions $x_1$ and $x_2$ are conjugated operators
\begin{equation}
[x_1(t_1), x_2(t_2)] = -i l_p L_{12} \, , 
\end{equation}
where $L_{12}$ is the distance between $x_1$ and $x_2$ and $l_p$ is the Planck length.
Thus the width of the position wave function describing their relative position increases with the separation like $\Delta x_1 \Delta x_2 = l_p L_{12}$. 

This uncertainty in relative position has impacts on gravity wave interferometers. Here this indeterminacy manifests as a maximum number of position eigenstates one can count in the frame of reference of the Michelson interferometer (like that at GEO600). If we assume that in any laboratory frame there is a maximum frequency $c/\lambda$ and a holographic information bound, so that on any light sheet (identified as the aforementioned Planck carrier wave) there is a minimum wavelength $\lambda$ in the transverse direction of the propagation direction of the light sheet, one gets the maximum number of position eigenstates that a interferometer can measure. 

Considering a 2+1--dimensional analogue of the situation we denote the coordinate axes by $x$, $y$ and $t$. There are two reference frames; first the light sheet frame which has wavefronts along $x$ and thus normal to $y$. The second frame is given by the $x-y$ plane spanned by both interferometer arms $L_x$ and $L_y$. Position measurements are carried out in the $x-y$ plane and the number of positions encoded in this plane is limited by the holographic principle.
Therefore, the maximum number of position eigenstates in the $x-y$ plane is given by the number of position eigenstates encoded on the light sheet frame (in this example it is the $x$--dimension) in units of the Planck length. 

Thus, this number is limited by $N< 2\, L_x/\lambda$. The interferometer arms of length $L=L_x=L_y$ measure the relative positions of bodies in the $x-y$--plane of the lab frame and the mean separation in each direction is denoted by $\Delta x$ then the number of position eigenstates is
\begin{equation}
\frac{L^2}{(\Delta x)^2}=N=\frac{2L}{\lambda},
\end{equation}
leading to
\begin{equation}
(\Delta x)^2 =\frac{\lambda L}{2}.
\end{equation}
This result is important as by considering the displacement $\Delta x$ one gets (by identifying $\lambda=l_{\rm P}$ ) $\Delta x \approx \sqrt{L l_{\rm P}}$ . In effect, the square root \textit{amplifies} the Planck-scale effect enormously compared with  the naive estimate that positions can only be located up to a fundamental length--scale of the order of $l_{\rm P}$. 

To sum up, in the 2+1 model the position information is encoded on the $x$--dimension in the light sheet frame in units of Planck-cells and the projection of this frame transverse to the direction of propagation generates the higher dimensional world. Furthermore, the main effect is that positions can be located on the light sheet frame with maximum precision of Planck-length whereas in the higher-dimensional world, due to the holographic principle, the location of positions is \textit{blurred}. 

The main result is that this error in localization is given by $\Delta x = \sqrt{L l_p}$ and can be interpreted as an indeterminacy of relative transverse positions. 

Therefore, Planck-scale precision cannot be achieved in the higher-dimensional world due to the holographic indeterminacy and it leads to an amplification of the intrinsic measurement error denoted here as fluctuations in length measurements.

It must be emphasized that in contrast to the aforementioned model of Ng holographic noise does not emerge here as perturbations of a metric. Nor is it devoted to any quantum gravity scenario or noncommutative geometry. Moreover, it is emphasized by Hogan that his approach respects Lorentz invariance and the weak equivalence principle on average. 

\section{GEO600 ``mystery noise'' and a possible explanation by holographic noise}\label{GEO600}

\subsection{GEO600 noise}

In the gravitational wave detector GEO600 an unexplained noise could not be identified while improving the sensitivity of this instrument. At present, the corresponding strain sensitivity is at the level of $10^{-21}$ lying in the frequency range of 500 Hz to about 1 kHz. Although a lot of experimental efforts are carried out, this noise cannot be eliminated at the moment. This opens up interesting questions about its origin including the speculation that it could be due to holographic noise.

\subsection{A possible explanation by holographic noise}

Applying the model of Hogan, after a time series of position measurements (and setting $\lambda=l_{\rm P}$) one gets the power spectral density $S\approx \Delta x^2 L/c$ which leads to a flat spectrum 
\begin{equation}
S=\frac{L^2 l_{\rm P}}{2c},
\end{equation}
where this corresponds to a magnitude of the amplitude spectral density of about $10^{-22}/\sqrt{\text{Hz}}$. This is just one order of magnitude smaller than the mysterious GEO600 noise. 

Although this proposal is appealing there remain several issues open. Because there is no connection to metrical fluctuations one can ask why the notion ``holographic noise'' applies for both models \cite{JackNg2003,{JackNg2002}} and \cite{Hogan_01_09}. Seen from a fundamental perspective, are both models describing the ``same'' holographic noise since they lead to different predictions ? Are there ``different'' kinds of holographic noise ? Furthermore it is claimed that there exists a derivation of this effect relying on a realization of M--theory as a matrix model \cite{MATRIX}.
Seen from a fundamental perspective this is highly debatable and needs more clarification. On the experimental side several assumptions are put forward (e.g. that the spot size of the lasers is minimized) which seems to be speculative and certainly needs more discussion. 

However, in the next sections we like to motivate a search for space--time fluctuations and especially holographic noise by means of matter--wave interferometers. This would supplement the measurements made at GEO600. In fact, if the ``mystery noise'' in GEO600 really has its origin in space--time fluctuations, then, owing to the universality of the coupling of geometry, this kind of noise should in principle show up in all physical phenomena and experimental devices.
 
\section{Modified quantum dynamics}

In the model \cite{GoeklueLaemmerzahl08} space--time is regarded as a fluctuating entity which consists of a classical fixed background on which Planck scale fluctuations are imposed $\eta_{\mu\nu}+ \delta g_{\mu\nu}$. In the low--energy limit they appear as classical fluctuations of space--time and we model this as perturbations of the metric up to second order 
\begin{eqnarray}
g_{\mu\nu}(\mathbf{x},t)& = & \eta_{\mu\nu}+ h_{\mu\nu}(\mathbf{x},t),  \\ \label{modmetric1}
g^{\mu\nu}(\mathbf{x},t)& = &\eta^{\mu\nu} - h^{\mu\nu}(\mathbf{x},t) + \tilde{h}^{\mu\nu}(\mathbf{x},t), \label{modmetric1a}
\end{eqnarray}
where $|h_{\mu\nu}| \ll 1$, greek indices run from 0 to 3 and latin indices from 1 to 3.
The second order perturbations are given by $\tilde{h}^{\mu\nu}=\eta_{\kappa\lambda}h^{\mu\kappa}h^{\lambda\nu}$ where indices are raised and lowered with $\eta_{\mu\nu}$ and $\eta^{\mu\nu}$.

We consider a scalar field $\phi(x)$ non--minimally coupled to gravity as given by the action
\begin{equation}
S = \frac{1}{2}\int d^4x \sqrt{g} \left(g^{\mu\nu} \partial_\mu\phi(x)^* \partial_\nu\phi(x) - \left(\frac{m^2c^2}{\hbar^2}+\xi R(x)\right) \phi(x)^* \phi(x)\right) \, ,
\end{equation}
where $g=-{\rm det} g_{\mu\nu}$ is the determinant of the metric and $x=(\mathbf{x},t)$. Here $R(x)$ is the Ricci scalar and $\xi$ a numerical factor. 

There are three values of $\xi$ which are of particular interest \cite{BirDav1982}: (i) $\xi=0$ is the minimally coupled case, (ii) $\xi=\frac{1}{6}$ is required by  conformal coupling and can also be derived from the requirement that the equivalence principle is valid for the propagation of scalar waves in a curved spacetime \cite{SonegoFaraoni93}. (iii) $\xi = \frac{1}{4}$ originates from the squaring the Dirac equation. 

Variation of the action with respect to $\phi(x)$ yields the Klein--Gordon equation non--minimally coupled to the gravitational field
\begin{eqnarray}
g^{\mu\nu}D_{\mu}\partial_{\nu}\phi(x)-\left(\frac{m^2c^2}{\hbar^2}+\xi R(x)\right)\phi(x)=0,
\end{eqnarray}
where $D_{\mu}$ is the covariant derivative based on the metric $g_{\mu\nu}$.
\subsection{Non--relativistic limit}
Because of the fact that in atomic interferometers typical velocities are far below the speed of light $c$ we only need the Schr\"odinger equation without relativistic corrections. For this, we have to calculate the non--relativistic approximation of the modified Klein--Gordon equation by expanding the wavefunction $\phi(\mathbf{x},t) = e^{iS(\mathbf{x},t)/\hbar}$ in powers of $c^2$
\begin{equation}
S(\mathbf{x},t)=S_0(\mathbf{x},t) c^2 + S_1(\mathbf{x},t) + S_2(\mathbf{x},t) c^{-2} + \ldots , \label{expansion1}
\end{equation}
according to the scheme worked out by Kiefer and Singh \cite{KieferSinghPRD1991}.
We give now a sketch of the calculations which were done in \cite{GoeklueLaemmerzahl08}. One performs a step-by-step calculation to orders $c^4$ and $c^2$ and inserts the expressions obtained for the functions $S_i$ into the equation of motion to order $c^0$, where non-hermitian terms appear in the Hamiltonian. In order to get rid of them one can choose a modified scalar product in curved space or alternatively keep the usual `flat' euclidean scalar product and transform the operators and the wavefunction accordingly \cite{LamPhysLetA1995}.
We proceed with the latter possibility and finally arrive at
\begin{eqnarray}
i \hbar \partial_t \psi&=& -(^{(3)}g)^{1/4}\frac{\hbar^2}{2m} \Delta_{{\rm cov}}\left((^{(3)}g)^{-1/4}\psi\right)-\frac{\hbar^2}{2m}\xi R(x)\psi + \frac{m}{2} g^{00}\psi\nonumber \\
& & +\frac{1}{2}\left\lbrace i \hbar \partial_i, g^{i0}\right\rbrace \psi\label{hamtrans}.
\end{eqnarray}
Here $\Delta_{{\rm cov}}$ represents the Laplace--Beltrami operator, $\lbrace \cdot , \cdot \rbrace$ is the anticommutator, $^{(3)}g$ is the determinant of the 3--metric $g_{ij}$ and $g^{00}$ includes the Newtonian potential $U(\mathbf{x})$. The Hamiltonian is manifest hermitian with respect to the chosen `flat' scalar product. 

\subsection{The effective Schr\"odinger equation}

It is assumed that the particle described by the modified Schr\"odinger equation has its own finite spatial resolution scale. Consequently, only an averaged influence of space--time fluctuations can be detected. This is quantified by means of the spatial average $\langle \cdots \rangle_{V}$ of the modified Hamiltonian leading to an effective Hamiltonian
\begin{equation}
H =-\frac{\hbar^2}{2m}\left(\left(\delta^{ij}+\alpha^{ij}(t)\right)\partial_i \partial_j+ \xi \left<R\right>_{V}\right) -m U(\mathbf{x})  \, . \label{effhamiltonian}
\end{equation}
where the tensorial function $\alpha^{ij}(t)$ consists of the spatial average of squares of the metrical fluctuations.

The tensorial function $\alpha^{ij}(t)$ and the Ricci scalar can be split into a positive definite time--average part and a fluctuating part according to
\begin{equation}\label{alphasplit}
\alpha^{ij}(t) = \tilde\alpha^{ij} + \gamma^{ij}(t)  \quad
{\rm and \, }\left<R\right>_{V}(t) = \tilde{r} + r(t)
\end{equation}
with 
\begin{equation}
\langle \gamma^{ij}(t) \rangle_T = 0 \quad {\rm and \,} \left< r(t) \right >_T = 0.
\end{equation}

Finally, this leads to an effective Schr\"odinger equation 
\begin{equation}
 i\hbar\partial_t \psi(\mathbf{x},t)=-\frac{\hbar^2}{2m}\left(\delta^{ij}+\tilde{\alpha}^{ij}\right)\partial_i \partial_j \psi(\mathbf{x},t) -\frac{\hbar^2}{2m}\tilde{r}\psi(\mathbf{x},t) - m U(\mathbf{x})\psi(\mathbf{x},t) \, , \label{effschroed}
\end{equation}
where the fluctuating phase $\gamma(t)+r(t)$ was absorbed into the wave function which is still denoted by the same symbol.

We will show in the following that this Schr\"odinger equation leads to an apparent violation of the weak equivalence principle \cite{GoeklueLaemmerzahl08}, to a modified dynamics of wavepackets \cite{BreuerGoeklueLaemml09} and to a decay of coherences in energy representation \cite{EGCLWavepacket2009}.
Since  $\tilde{r}$ is constant over the averaging scale it just represents a constant shift of energy what can be omitted. Therefore we will operate in the following with the effective Hamiltonian 
\begin{equation}
H= -\frac{\hbar^2}{2m}\left(\delta^{ij}+\tilde{\alpha}^{ij}\right)\partial_i \partial_j  - m U(\mathbf{x}). \label{effhamiltonian1}
\end{equation}

\section{An apparent violation of WEP}\label{WEPVIOLATION}
It will be shown now that the fluctuation tensor $\tilde{\alpha}^{ij}$ leads to a renormalized inertial mass implying an apparent breakdown of the Weak Equivalence Principle (WEP).
We identify the mass term in the Hamiltonian (\ref{effhamiltonian1}) as a renormalized tensorial inertial mass by virtue of 
\begin{equation}
(\tilde{m}^{ij})^{-1}\equiv \frac{1}{m}\left(\delta ^{ij}+ \tilde{\alpha}^{ij}\right),
\end{equation}
what can alternatively be written as 
\begin{equation}
\tilde{m}_i\equiv m \cdot (1+ \alpha_p^i)^{-1},
\end{equation}
where $\tilde{\alpha}^{ij}={\rm diag }\left[\alpha_p^1, \alpha_ p^2, \alpha_p^3\right]$ and $\alpha_p^i=\delta^{ij}\left<\tilde{h}_{(0)}^{ii}\right>_{V_p}$. The index $p$ stands for ''particle'' characterized by its spatial resolution scale $\lambda_p=V_p^{1/3}$.
This kind of anomalous inertial mass tensor was extensively studied by Haugan in connection with the principle of equivalence (EP) \cite{Haugan1979AnnPhys}. There, in contrast to the above result, this quantity is postulated in order to model violations of EP and appears in the energy function of a composite body subject to an external gravitational potential.

All space--time dependent fluctuation properties of the inertial mass are now encoded in the quantities $\alpha^i$ which are given by the averaged quadratic perturbations $\sigma^2_{ii}$.

When we identify $ \tilde{m}_j=m_{\rm i}$ and $ m=m_{\rm g}$ as inertial and gravitational mass giving for the ratio of both masses 
\begin{equation}
\left(\frac{m_{\rm g}}{m_{\rm i}}\right)^{i}_p = 1 + \alpha^{i}_p \, ,
\end{equation}
we can conclude that the metric fluctuations utilized in our model lead to an apparent violation of the weak equivalence principle. 

\subsection{Holographic noise and other noise scenarios}
In analogy to the analysis of fluctuations as function of time carried through in \cite{Radeka1988} we can associate a spectral density $S(\mathbf{k})$ with the variance, 
\begin{equation}
(\sigma^2)^{ij} = \frac{1}{V_p} \int_{(1/\lambda_p)^3} d^3k  (S^2(\mathbf{k},t))^{ij}= \frac{1}{V_p} \int_{V_p} d^3x \tilde{h}^{ij}(\mathbf{x},t) \, ,
\end{equation}
which allows us - in principle - to calculate the metric fluctuation components $\tilde{\alpha}^{ij}$. 

As a result we then obtain $(\sigma^2)^{ij} \approx \left(l_{\rm Pl}/\lambda_p\right)^{\beta}a^{ij}$, where $a^{ij}$ is a diagonal tensor of order $\mathcal{O}(1)$.
The special model \cite{JackNg2003,JackNg2002} which we discussed in section \ref{MODHOL} can now be applied.  According to this, the outcome of any experiment measuring length intervals is influenced by spacetime fluctuations such that perturbations of this quantity scales like $\left(l_{\rm Pl}/l\right)^{\beta}$, where $\beta\geq 0$. 
As pointed out in \cite{JackNg2003} common values for $\beta$ are $1,2/3,1/2$. 

We emphasize that by the inclusion of $\beta=2/3$ (corresponding to the fluctuations of the length $\delta l \approx (l_p^2 l)^{1/3}$ ) a holographic noise scenario is also included in our model of space--time fluctuations.\\
As a consequence, the modified kinetic term of the averaged Hamilton operator reads
\begin{equation}
H_{{\rm kin.}}=-\frac{\hbar^2}{2m}\left(\delta^{ij}+\left(\frac{l_{\rm Pl}}{\lambda_p}\right)^{\beta}a^{ij}\right)\partial_i \partial_j  \, . \label{flucscenario}
\end{equation}
This gives for our fluctuation scenario
\begin{equation}
\tilde{m}_{i}=\frac{m}{1+(l_{\rm Pl}/\lambda_p)^{\beta} a^{ii}}.
\end{equation}
In terms of the fluctuation scenario (\ref{flucscenario}) the violation factor $\alpha_p$ now yields
\begin{equation}
\alpha^i_p=\left(\frac{l_{\rm Pl}}{\lambda_p}\right)^{\beta}a^{ii} \, .
\end{equation}
The modified inertial mass $\tilde{m}_i$ is now dependent on the fluctuation scenario $\beta$ and on the type of particle via the characteristic resolution scale $\lambda_p$.
\subsection{Experimental consequences}

In order to obtain some estimates, we set the resolution scale to the Compton length $\lambda_{pj}=\hbar/(m_j c)$ being a characteristic property of every particle species. For Cesium this gives $\lambda_{\rm Cs}\approx 10^{-18}\;{\rm m}$ and for Hydrogen we get $\lambda_{\rm H}\approx 10^{-16}\;{\rm m}$. Assuming rotational invariance ($\alpha^i_p=\alpha_p $) we get
\begin{equation}
\alpha_{\rm Cs}\approx\left(\frac{10^{-35}} {10^{-18}}\right)^{\beta}=(10^{-17})^{\beta}\, ,  \quad \alpha_{\rm H}\approx\left(\frac{10^{-35}} {10^{-16}}\right)^{\beta}=(10^{-19})^{\beta}\, .
\end{equation}
For the calculation of the E\"otv\"os factor we need the accelerations, namely
\begin{equation}
a_{\rm Cs}=(1+\alpha_{\rm Cs})g=(1+(10^{-17})^{\beta})g, \quad a_{\rm H} = (1 + \alpha_{\rm H})g=(1+(10^{-19})^{\beta}) g.
\end{equation}
The E\"otv\"os factor reads  
\begin{eqnarray}
\eta = 2\frac{|a_{\rm Cs} - a_{\rm H}|}{|a_{\rm Cs} + a_{\rm H}|} 
\approx |\alpha_{\rm Cs} - \alpha_{\rm H}| \approx (10^{-17})^{\beta}.
\end{eqnarray}
In our model this yields
\begin{eqnarray}
\eta_{\beta=1}=10^{-17}, \quad \eta_{\beta=2/3}=10^{-12}, \quad \eta_{\beta=1/2}=10^{-9}.
\end{eqnarray}
Regarding the precision of current atom interferometers \cite{PetChungChu1999} this would rule out the random--walk scenario $\beta=1/2$, but one has to take into account that our scenario is quite optimistic, regarding the amplitude of fluctuations estimated by the spatial resolution scale $\lambda_p=\lambda_C$.  

However, it is worth to mention that the holographic noise scenario $\beta=2/3$ may be tested by high--precision atomic interferometry in the near future as it seems not to be too far off from experimental possibilities. 

\section{Spreading of wavepackets}\label{WAVEPACKETS}
\subsection{Approximation of the kinetic term}
In the preceding section we analyzed the modified kinetic term in (\ref{effhamiltonian}) and its implication on the weak equivalence principle. By using spatial averaging and a modified time evolution operator we showed that the effect of non-minimal coupling, given by the Ricci scalar, leads to a fluctuating phase of the wavefunction. However, now we will start again with the full Hamiltonian of (\ref{hamtrans}) and analyze the impact of the modified kinetic term on wavepacket spreading by deriving a modified master equation \cite{EGCLWavepacket2009}.

All contributions of the kinetic part as well as those terms in the anticommutator of equation (\ref{hamtrans}) which do not include derivatives can be summed up together with the Ricci scalar in a stochastic scalar interaction term $V(\mathbf{x},t)$ appearing in the Schr\"odinger equation
\begin{eqnarray}
i \hbar \partial_t \psi&=& -\frac{\hbar^2}{2m} \left( g^{ij}(\mathbf{x},t)\partial_i\partial_j+\partial_i g^{ij}\partial_j\right) \psi + i\hbar g^{i0}\partial_i \psi + V(\mathbf{x},t)\psi \label{schroed1},
\end{eqnarray} 
where for simplicity we omitted the Newtonian potential $g^{00}$. Our conclusions will not be affected by taking a Newtonian gravitational field in to account. 

Since we employ spacetime fluctuations acting on a spacetime scale being much smaller than typical length- and timescales on which the wavefunction $\psi(\mathbf{x},t)$ varies we approximate the Hamiltonian in such a way that terms containing first- and second derivatives of the fluctuating quantities ($h^{ij}$ and the trace $h$) dominate in the Schr\"odinger equation (\ref{schroed1}) allowing us to neglect the other contributions. 

In addition, we note that in the following derivation of the averaged master equation only second moments of $V$ will appear, see Eq. (\ref{corr3}). Therefore we can neglect second order terms already on the level of the Schr\"odinger equation leading to
\begin{equation}
 i \hbar \partial_t \psi=-\frac{\hbar^2}{2m}\Delta \psi +V(\mathbf{x},t) \psi \label{schroed2},
\end{equation}
where $\Delta$ is the Laplace operator in Cartesian coordinates and where the random scalar interaction $V(\mathbf{x},t) $ to first order now reads
\begin{eqnarray}
V(\mathbf{x},t)& = & -\frac{\hbar^2}{2m}\left( \xi R-\frac{1}{4}\Delta h\right), \label{randpot1}
\end{eqnarray}
where $h$ is the first-order spatial part of the trace $\text{tr}(g_{\mu\nu})$.

Compared to the second term in the fluctuating potential (\ref{randpot1}) the non-minimal coupling term $\xi R$ has the same structure and the same order of magnitude. Thus this contribution to the Schr\"odinger Hamiltonian and to the analysis of spacetime fluctuations cannot be neglected in the context of our model and must be taken into consideration in the following calculations.

\subsection{Stochastic properties}

We have to specify the statistical properties of the fluctuating potential $V(\mathbf{x},t)$ in order to derive an effective master equation. 
Owing to the fluctuating metric this term will be interpreted as a Gaussian random function. 
We will now perform the calculations for one dimension as the effective master equation will be derived for this specific case and from now on the spatial coordinate is denoted by $x$.
This yields for the fluctuating potential
\begin{eqnarray}
 V(x,t) & = & \frac{\hbar^2}{8m} \partial_x^2 h_{11}(x,t).
\end{eqnarray}
As the quantities of interest are the metric perturbation terms, we choose the following statistical conditions
\begin{eqnarray}
\left<V(x,t)\right> & = &0 \label{corr1}\\
\left<V(x,t)V(x',t')\right> & = & V_0^2 \delta(t-t') g(x-x') \label{corr3}.
\end{eqnarray}
We introduced $g(x-x')=\partial_x^2 \partial_{x'}^2 C(x-x')$, where $C(x-x')$ is a spatial correlation function and  defined $V_0=\frac{\hbar^2}{8 m}h_0$, while $h_0$ is the strength of the fluctuations.
It has been shown by Heinrichs \cite{Heinrichs92, {Heinrichs96}} that the dynamics of wavepackets are unaffected by temporal correlations of the random function $h(x,t)$ if they are sufficiently small.
As we are concerning Planck-scale effects (Planck time $\tau_p$ and Planck length $l_p$) we can adopt these arguments. Therefore we have chosen a $\delta$--correlation with respect to time. However, at the moment the spatial correlation function $C(x-x')$ will be left unspecified.

As this is a model for spacetime fluctuations and because of the lack of a complete understanding of the microscopic structure of spacetime one has the freedom to characterize the statistical properties of the terms $h$ with a variety of stochastic models given here by the correlator $C(x-x')$. 

\subsection{Effective master equation}

Most physical quantities of interest -- in our case the mean square displacement -- can be conveniently calculated in terms of the reduced density matrix $\left<\rho(x',x,t)\right>$ and its evolution equation. For the sake of simplicity we restrict all the calculations to the one--dimensional case. We adopt the calculation techniques of  \cite{Jayannavar82, {Jayannavar93}} which are used in conjunction with quantum diffusion of particles in dynamical disordered media. 

We define the density operator 
\begin{equation}
 \rho(x',x,t)=\psi^*(x',t)\psi(x,t)
\end{equation}
and set up the master equation according to
\begin{eqnarray}
\partial_t \rho(x',x,t)& = & \frac{i \hbar}{2m}\left(\frac{\partial^2}{\partial x^2}-\frac{\partial^2}{\partial x'^2}\right)\rho(x',x,t)+\frac{i}{\hbar}\left(V(x',t)-V(x,t)\right)\rho(x',x,t). \label{mastereq}
\end{eqnarray}

The effective equation of motion can now be derived by applying the average over the fluctuations. 
Terms like $\left<V(x,t)\rho(x',x,t)\right>$ do not factorize in general as the density operator $\rho(x',x,t)$ is a functional of the fluctuation potential, hence $\rho[V]$. Because we have Gaussian fluctuations, we can apply the Novikov theorem \cite{Novikov64}, which leads in our case to
\begin{eqnarray}
 \left<V(x,t)\rho(x',x,t)\right> & = & \int dt'' \int dx'' \delta(t-t'') g(x-x'')\left<\frac{\delta  \rho(x',x,t)}{\delta V(x'',t'')}\right> \label{novikov1}.
\end{eqnarray}

The functional derivatives are calculated using the formal solution of the master equation. 
Inserting this into the averaged Equation (\ref{mastereq}) we obtain the effective master equation for the reduced density matrix $\left<\rho(x',x,t)\right>$
\begin{eqnarray}
\frac{\partial }{\partial t}\left<\rho(x',x,t)\right>  & = & -\frac{i \hbar}{2m}\left(\frac{\partial^2}{\partial x'^2}-\frac{\partial^2}{\partial x^2}\right)\left<\rho(x',x,t)\right> \nonumber \\
& & - \frac{V_0^2}{\hbar^2}\left(g(0)-g(x-x')\right)\left<\rho(x',x,t)\right>. \label{modmastereq}
\end{eqnarray}
This is the modified effective master equation subject to the following discussions.
\subsection{Modified wavepacket spreading}
It is convenient to calculate the solution by means of Laplace- and Fouriertransforms. Hence, we convert the partial differential equation (\ref{modmastereq}) to an ordinary, inhomogeneous first-order differential equation.  
For practical reasons we introduce new coordinates according to $X=x+x'$ and $Y=x-x'$ and after having applied both transformations we get
\begin{equation}
 \partial_Y R(K,Y,s) - \frac{m}{2\hbar K} \left(s+\frac{V_0^2}{\hbar^2}\left(g(0)-g(Y)\right)\right)R(K,Y,s)=- \frac{m}{\hbar K}R(K,Y,t=0) \, ,
\end{equation}
where is the $s$ is the Laplace-variable, while $R(K,Y,t=0)$ is the Fourier transform of $\left<\rho(X,Y,t=0)\right>$ and represents now the inhomogeneous part of the ordinary differential equation.

This equation can easily be integrated and yields
\begin{eqnarray}
R(K,Y,s)& = & \exp{\left(\frac{m s Y}{2\hbar K}+G(Y) \right)}R(K,Y_0,s)-\exp{\left(\frac{m s Y}{2\hbar K}+G(Y) \right)}\times \label{solution1} \\
& & \times\frac{m}{\hbar K}\int_0^Y dY' \exp{\left(-\frac{msY'}{2\hbar K}\right)} R(K,Y',t=0) \exp{\left(-\frac{mG(Y')}{2 \hbar K}\right)}, \nonumber 
\end{eqnarray}
where the term $R(K,Y_0,s)=R(K,Y=0,s)$ represents the initial value of the solution $R(K,Y,s)$ and we defined $G(Y)=\frac{V_0^2}{\hbar^2}\int^Y_0 dY' \left(g(0)-g(Y')\right)$.
\\
The mean squared displacement of the particle can be expressed in Fourier space as 
\begin{eqnarray}
\sigma_x^2(t) =-\frac{1}{8}\frac{\partial^2}{\partial K^2}R(K,Y_0,t)\Big|_{K=0}.\label{msdisplacement}
\end{eqnarray}
Therefore the quantity of interest which has to be calculated is the initial value of $R$ at $Y=0$, corresponding to the diagonal element of the density matrix $\rho(x',x,t)$. 

The solution $R(K,Y,s)$ is a  quadratically integrable function and must vanish for $Y\rightarrow \infty$. Therefore the r.h.s. of equation (\ref{solution1}) must be zero in this limit leading to the Laplace-transformed quantity
\begin{equation}
 R(K,Y_0,s)=2 \int_0^{\infty} d\tau  \exp{\left(-s\tau\right)} R(K,\frac{2 \hbar K}{m}\tau,t=0) \exp{\left(-G(\tau)\right) },
\end{equation}
where we applied the substitution $\tau=\frac{m}{2\hbar K}Y$.
Inverting this equation gives
\begin{equation}
R(K,Y_0,\tau)= 2 \, R(K,\frac{2 \hbar |K|}{m}\tau,t=0)\exp{\left(-\frac{V_0^2}{\hbar^2}\int^{\tau}_{0} d\tau'  \left[g(0)-g\left(\frac{2 \hbar |K|}{m}\tau'\right)\right]\right)}
\end{equation}
\\
and the mean squared displacement (\ref{msdisplacement}) reads
\begin{equation}
 \sigma_x^2(t) = \frac{1}{4}\,{{\rm e}^{-G (Y)}}R_0(Y) \left(
\partial_k ^{2}G(Y) - \left(\partial_k G(Y)  \right) ^{2} +2\, \partial_k G(Y) \partial_k \ln{R_0(Y)}-\partial^2_k \ln{R_0(Y)}-(\partial_k \ln{R_0(Y)})^2  \right)\Big|_{K=0}, \label{msdispplacementgeneral}
\end{equation}
where $Y=2 \frac{\hbar |K|}{m}\tau$ and $R_0(Y)=R(K,\frac{2 \hbar |K|}{m}\tau,t=0)$.
\\

Note that the unperturbed solution is given by the last two logarithmic terms, where for $K=0$ we get $\exp{(-G (Y))}R_0(Y)=1$, if the initial state $R_0$ is given by a Gaussian wavepacket. The information about the fluctuation scenario is given by the correlation function appearing in $G(Y)$. In our case we regard even correlation functions $G(Y)=G(-Y)$ excluding a preferred direction in space.

We can now calculate the mean squared displacement of a wavepacket and choose a Gaussian correlator
\begin{equation}
 C(Y)=\frac{1}{\sqrt{2\pi}a}\exp{\left(-\frac{Y^2}{a^2}\right)} \label{corrspat1},
\end{equation}
where $a$ represents a finite correlation length and we choose for the initial conditions $R(K,Y,t=0)$ a Gaussian wavepacket.
Consequently, the mean squared displacement reads
\begin{eqnarray}
 \sigma_x^2(t) = \sigma^2_x(0)+ \frac{\hbar^2}{4m^2\sigma^2_x(0)}t^2 + \frac{5 V_0^2}{\sqrt{2\pi}m^2 a^7} t^3. \label{msd1}
\end{eqnarray}
We note that the first two terms correspond to the free unperturbed evolution whereas the last term accounts for superdiffusive behavior. \footnote{The nomenclature is at follows: For the mean-squared displacement $\left<x^2\right>\propto t^{\nu}$ the exponent $\nu=1$ renders diffusive motion, $\nu=2$ is a ballistic motion and $\nu=3$ denotes superdiffusion.}. Asymptotically, the wavepacket dynamic is dominated by the cubic term, namely
\begin{equation}
 \sigma^2_x(t) = \sigma^2_x(0) +\frac{5 V_0^2}{\sqrt{2\pi}m^2 a^7} t^3 , \qquad \text{for } t\rightarrow \infty.\label{msd3}
\end{equation}
We infer from this fact that space--time fluctuations lead to superdiffusive behavior of wavepackets and resembles wavepacket spreading in dynamically disordered media \cite{Jayannavar93}. 
However, it must be emphasized that the magnitude of this effect in our model is small since the amplitude $V_0$ exhibits Planck-scale contributions 
\subsection{Connection to holographic noise}
In the holographic scenario of \cite{JackNg2003,{JackNg2002}} the fluctuations of the metric $\delta g_{\mu\nu} \geq (l_p / l)^{2/3} a_{\mu\nu}$
can be translated to a power spectral density (PSD) according to 
\begin{equation}
S(f)=f^{-5/6}(c l^2_p)^{1/3},
\end{equation}
where $f$ is the frequency and $c$ is the speed of light. 
In terms of a PSD in momentum space this yields
\begin{equation}
S(k)=k^{-5/6} l_p^{2/3},
\end{equation}
where $k$ is the wavenumber. It can be readily checked that this expression leads to $\delta l = (l\, l_p^2)^{1/3}$ by calculating the variance $(\delta l)^2=\int^{k_{max.}}_{1/l} dk \left(S(k)\right)^2$, where $k_{max.}$ is subject to a cut-off given by the Planck length $l_p$ and the length $l$ is given by the experiment.
This shows that - by means of a correlation function - a finite, nonvanishing correlation length is involved where the correlator scales with $ l_p^{2/3}$ modifying the wavepacket evolution accordingly. Therefore the $l_p^{2/3}$-dependence could be utilized to distinguish the holographic noise scenario for wavepacket spreading from other fluctuations scenarios.

However, in order to test these scenarios with high precision we suggest to use cold atoms and long time of flights. In this context it is feasible that test should be carried out in microgravity environments to facilitate the free evolution of wavepackets. This is realized for example by the  Bose--Einstein--Condensate of the QUANTUS project \cite{QUANTUS06} where free evolution times up to one second are reached. Furthermore the upcoming PRIMUS project will provide a matter--wave interferometer in free fall which also could be used to perform high--precision measurements of wavepacket spreading. 
\section{Decoherence due to space--time fluctuations}\label{SEC-QMEQ}

\subsection{Effective master equation}
Having redefined the inertial mass of the particle as described in the preceding section \ref{WEPVIOLATION} we are left with an effective Schr\"odinger equation of the form 
\begin{equation}
\partial_t  |\tilde{\psi}(t)\rangle = -\frac{i}{\hbar} \left( H_0 + H_p(t) \right)
 |\tilde{\psi}(t)\rangle, \label{EffectiveSchroedinger}
\end{equation}
where, however, only the fluctuating part $\gamma^{ij}(t)$ enters
the Hamiltonian $H_p(t)$ and we write
\begin{equation}
 H_p(t) = \frac{1}{2m} \gamma^{ij}(t) p_i p_j,  \qquad H_0 = \frac{\mathbf{p}^2}{2m} - mU, \label{DEFS}
\end{equation}
where $p_i = -i\hbar \partial_i$ and $H_0$ is defined with an appropriately renormalized inertial mass in the kinetic term.
We start by transforming to the interaction picture and obtain the Schr\"odinger equation
\begin{equation}
 \frac{d}{dt} \tilde{\psi}(t) = -\frac{i}{\hbar} \tilde{H}_p(t)
 |\tilde{\psi}(t)\rangle, \label{SSE}
\end{equation}
where the tilde labels quantities in the interaction picture. 
Equation ~(\ref{SSE}) represents a stochastic Schr\"odinger equation (SSE) involving a random Hamiltonian
$\tilde{H}_p(t)$ with zero average, $\langle \tilde{H}_p(t) \rangle = 0$. For a given realization of the random process
$\gamma^{ij}(t)$ the corresponding solution of the SSE represents a pure state with the density matrix $\tilde{R}(t)$
satisfying the von Neumann equation
\begin{equation} \label{StochNeumann}
 \frac{d}{dt}\tilde{R}(t) = -\frac{i}{\hbar}
 \left[\tilde{H}_p(t),\tilde{R}(t)\right] \equiv
 {\mathcal{L}}(t)\tilde{R}(t),
\end{equation}
where ${\mathcal{L}}(t)$ denotes the Liouville superoperator.
However, if we consider the average over the fluctuations of the $\gamma^{ij}(t)$ the resulting density matrix of the Schr\"odinger particle,
\begin{equation} \label{DENSITY}
 \tilde{\rho}(t) = \left\langle \tilde{R}(t) \right\rangle
\end{equation}
generally represents a mixed quantum state. Thus, when considering averages, the dynamics given by the SSE transforms pure states into mixtures and leads to dissipation and decoherence processes. Consequently, the
time-evolution of $\tilde{\rho}(t)$ must be described through a dissipative quantum dynamical map that preserves the Hermiticity, the trace and the positivity of the density matrix \cite{OUP}.

We derive from the linear stochastic differential equation (\ref{StochNeumann}) an equation of motion for the average (\ref{DENSITY}) by the cumulant expansion method \cite{KAMPEN12}.
To second order in the strength of the fluctuations it yields the equation of motion
\begin{eqnarray} \label{MASTEREQ}
 \frac{d}{dt}\tilde{\rho}(t) & = & -\frac{1}{\hbar^2} \int_0^t dt_1
 \left\langle \left[\tilde{H}_p(t),\left[
 \tilde{H}_p(t_1),\tilde{\rho}(t)\right]\right]\right\rangle.
\end{eqnarray}
This is the desired quantum master equation for the density matrix of the Schr\"odinger particle, representing a local first-order differential equation with time-dependent coefficients.

\subsection{Markovian master equation}

To proceed further we have to specify the stochastic properties of the random quantities $\gamma^{ij}(t)$. We consider a white noise scenario assuming that the fluctuations are isotropic,
\begin{equation} \label{ansatz}
\gamma^{ij}(t) = \sigma\delta_{ij} \xi(t).
\end{equation}
This assumption already singles out a certain reference frame, which can be identified with the frame in which the space averaging has been carried out \cite{GoeklueLaemmerzahl08}. In (\ref{ansatz}) the function $\xi(t)$ is taken to be a Gaussian white noise process with zero mean and a $\delta$-shaped auto-correlation function,
\begin{equation} \label{CORR}
\langle \xi(t) \rangle = 0, \qquad \langle \xi(t) \xi(t') \rangle
= \delta(t-t').
\end{equation}
Therefore, the quantity $\sigma^2$ has the dimension of time and we set  $\sigma^2 = \tau_c$.
Within the white noise limit the fluctuations thus appear as un-correlated on the time scale of the particle motion
and the contributions from the higher-order cumulants vanish.  Then the second-order master equation
(\ref{MASTEREQ}) becomes an exact equation \cite{KAMPEN3}. Using (\ref{ansatz}) we find
\begin{equation}
 \tilde{H}_p(t) = \hbar \tilde{V}(t) \xi(t),
\end{equation}
where
\begin{equation}
 \tilde{V}(t) = e^{iH_0t/\hbar} V e^{-iH_0t/\hbar}, \qquad
 V = \frac{\sqrt{\tau_c}}{\hbar} \frac{\textbf{p}^2}{2m}.\label{PERTOPERATOR}
\end{equation}
Substitution into the master equation (\ref{MASTEREQ}), employing (\ref{CORR}) and transforming back to the Schr\"odinger picture we finally arrive at the master equation
\begin{equation} \label{QMEQ}
 \frac{d}{dt}\rho(t) = -\frac{i}{\hbar}[H_0,\rho(t)] +
 {\mathcal{D}}(\rho(t)),
\end{equation}
where
\begin{equation} \label{DISS}
 {\mathcal{D}}(\rho(t)) = -\frac{1}{2}
 \left[V,\left[V,\rho(t)\right]\right]
 = V\rho(t)V-\frac{1}{2}\left\{V^2,\rho(t)\right\}.
\end{equation}
Equation (\ref{QMEQ}) represents a Markovian quantum master equation for the Schr\"odinger particle. The commutator term involving the free Hamiltonian $H_0$ describes the contribution from the
coherent motion, while the superoperator ${\mathcal{D}}(\rho)$, known as \textit{dissipator}, models all dissipative effects. We observe that the master equation is in Lindblad form and, thus, generates a completely positive quantum dynamical semigroup \cite{GORINI,LINDBLAD}.

\subsection{Estimation of decoherence times}
The quantum master equation (\ref{QMEQ}) can easily be solved in the momentum representation. To this end, we define the density matrix in the momentum representation $\left |\mathbf{p}\right>$ and with the help of the master equation we then find
\begin{eqnarray}
 \frac{d}{dt}\rho({\mathbf{p}},{\mathbf{p}}',t)
 & = & -\frac{i}{\hbar}\left[E({\mathbf{p}})-E({\mathbf{p}}')\right]
 \rho({\mathbf{p}},{\mathbf{p}}',t)\nonumber \\
 & & -\frac{\tau_c}{2\hbar^2}\left[E({\mathbf{p}})-E({\mathbf{p}}')\right]^2
 \rho({\mathbf{p}},{\mathbf{p}}',t),
\end{eqnarray}
where $E({\mathbf{p}})={\mathbf{p}}^2/2m$. This equation is
immediately solved to yield
\begin{equation}
 \rho({\mathbf{p}},{\mathbf{p}}',t) =
 \exp\left[ -\frac{i}{\hbar}\Delta E t - \frac{(\Delta E)^2\tau_c}{2\hbar^2} t \right]
 \rho({\mathbf{p}},{\mathbf{p}}',0),
\end{equation}
where $\Delta E=E({\mathbf{p}})-E({\mathbf{p}}')$. We see that the diagonals of the density matrix in the momentum representation are constant  in time.
On the other hand, the coherences corresponding to different energies decay exponentially with the rate $(\Delta E)^2\tau_c/2\hbar^2$. Therefore an associated decoherence time $\tau_D$ is given by
\begin{equation} \label{tau-d}
 \tau_D = \frac{2\hbar^2}{(\Delta E)^2\tau_c} =
 2\left(\frac{\hbar}{\Delta E\cdot\tau_c}\right)^2 \tau_c.
\end{equation}
We see that the dissipator ${\mathcal{D}}(\rho)$ of the master equation leads to a decay of the coherences of superpositions of energy eigenstates with different energies, resulting in an effective
dynamical localization in energy space. 
This feature of the master equation is due to the fact that the fluctuating quantities $\gamma^{ij}(t)$ couple to the components of the momentum operator.

In our model the different possible realizations of the wavefunction due to the stochastic fluctuations of spacetime are
accounted for by the ensemble average introduced in equation (\ref{DENSITY}). In a matter wave experiment this leads to a reduced contrast of the interference fringes when both atom beams are recombined. When the travelling time of both beams is of the order of the decoherence time $\tau_D$ the interference fringes get
smeared out significantly indicating the partial destruction of quantum coherence.

Let us identify the time scale $\tau_c$ that characterizes the strength of the fluctuations with the Planck time $t_p$  and we set $\tau_c =t_p = l_p/c$ with the Planck length $l_p$. The expression (\ref{tau-d}) then yields the estimate
\begin{equation}
\tau_D \approx \frac{10^{13}{\mathrm{s}}}{(\Delta
E/{\mathrm{eV}})^2}.
\end{equation}
We observe that due to the quadratic behavior the decoherence time depends strongly on the scale of the energy difference $\Delta E$. For example, $\Delta E = 1{\mathrm{eV}}$ leads to a decoherence time of the order of 300,000 years ($10^{13}$ seconds), while $\Delta E = 1{\mathrm{MeV}}$ leads to a decoherence time of the order of $10$ seconds.
Therefore for typical matter--wave experiments the detection of decoherence effects seems not to be feasible and is far beyond any experimental relevance. 

However, this result does not rule out in general the experimental decoherence caused by metric fluctuations if one considers, for example, composite quantum objects whose states can be extremely sensitive to environmental noise.

In principle the white--noise scenario (\ref{CORR}) could be generalized to colored noise including holographic noise. This would lead to additional non--Markovian terms in the master equation (\ref{QMEQ}) and we expect, however, that
such terms would manifest only as small corrections to the equation for the decoherence time $\tau_D$.

\section{Concluding remarks}
We have shown that metric fluctuations affect quantum dynamics where we have left open which noise scenario is realized due to the lack of a final theory of quantum gravity. One candidate is the holographic noise scenario and we have shown that this leads to specific signatures in the quantities of interest, which could be detected by experiments. These are the E\"otv\"os factor, the mean--squared displacement of wave--packets and, when considering non--Markovian dynamics, the decoherence time of quantum systems. 
Since the topic of this workshop was to investigate the possibility of the detection of gravitational waves by means of atom interferometers we like to add some remarks about how our work is related to this. The detection of gravitational waves needs exceptional sensitivity and since space--time fluctuations manifest as tiny effects these improved devices would be feasible to detect metric fluctuations or to rule out our scenario. Furthermore, the possibility that GEO600 may have detected holographic noise should give motivation to check this with an independent method. Since gravity couples to all physical systems this effect may show up in high--precision atom interferometers which are dedicated to the detection of gravitational waves.

\section*{Acknowledgements}

We like to thank K. Danzmann, D. Giulini, C. Hogan, and S. Hild for helpful  discussions. EG gratefully acknowledges the support by the German Research Foundation (DFG) and the Centre for Quantum Engineering and Space--Time Research (QUEST), and CL the support by the German Aerospace Center (DLR) grant no. 50WM0534.

\end{document}